\documentclass[conference]{IEEEtran}
\IEEEoverridecommandlockouts
\usepackage{cite}
\usepackage{amsmath,amssymb,amsfonts}
\usepackage{graphicx}
\usepackage{textcomp}
\usepackage{xcolor}
\usepackage{url}
\usepackage{etoolbox}
\def\BibTeX{{\rm B\kern-.05em{\sc i\kern-.025em b}\kern-.08em
    T\kern-.1667em\lower.7ex\hbox{E}\kern-.125emX}}

\newcommand{\RR}{\mathbb{R}}
\newcommand{\NN}{\mathbb{N}}
\newcommand{\PP}{\mathbb{P}}
\newcommand{\EE}{\mathbb{E}}
\newcommand{\SINR}{\mathrm{SINR}}
\newcommand{\cm}{\mathrm{c}}
\newcommand{\sn}{\mathrm{s}}
\newcommand{\dd}{\mathrm{d}}
\newcommand{\tr}{\operatorname{tr}}
\newcommand{\Cov}{\operatorname{Cov}}

\newcommand{\ol}[1]{\overline{#1}}
\newcommand{\Xt}{\tilde X}

\newcommand{\figslot}[4]{%
  \IfFileExists{#1.pdf}{\includegraphics[width=#2,height=#3,keepaspectratio]{#1.pdf}}{%
  \IfFileExists{#1.png}{\includegraphics[width=#2,height=#3,keepaspectratio]{#1.png}}{%
  \fbox{\parbox[c][#3][c]{\dimexpr#2-2\fboxsep-2\fboxrule\relax}{%
    \centering\scriptsize\textcolor{red}{\texttt{\detokenize{#1}}\\[2pt]#4}}}}}}

\newtheorem{theorem}{Theorem}
\newtheorem{lemma}{Lemma}
\newtheorem{proposition}{Proposition}

\newtheorem{definition}{Definition}
\newtheorem{assumption}{Assumption}
\newtheorem{remark}{Remark}
\newcommand{\blockspace}[1]{%
  \BeforeBeginEnvironment{#1}{\par\addvspace{\smallskipamount}}%
  \AfterEndEnvironment{#1}{\par\addvspace{\smallskipamount}}}
\forcsvlist{\blockspace}{theorem,lemma,proposition,corollary,definition,assumption,remark,IEEEproof}
\begin{document}

\title{System-level performance in massive ISAC networks}

\author{\IEEEauthorblockN{Emanuele Mengoli}
\IEEEauthorblockA{\textit{T\'el\'ecom Paris}\\
\textit{INRIA Paris}\\
Paris, France\\
emanuele.mengoli@inria.fr}
\and
\IEEEauthorblockN{Nahuel Soprano-Loto}
\IEEEauthorblockA{
\textit{INRIA Paris}\\
Paris, France \\
nahuel.soprano-loto@inria.fr}
}

\maketitle

\begin{abstract}
We propose a dynamic stochastic geometry framework to model and evaluate large-scale integrated sensing and communication (ISAC) networks. Base stations are distributed according to a homogeneous Poisson point process, while user equipments and sensing objects undergo localised random motion around their serving base station. Full spectrum reuse with a common waveform generates mutual interference that couples the two functionalities. Communication performance is characterised through downlink queues whose service rates are governed by the signal-to-interference-plus-noise ratio (SINR) via the Shannon capacity. Sensing performance is captured by a Kalman filter (KF) whose measurement noise covariance depends on the sensing SINR, yielding a non-classical state-dependent filtering problem. We define system-level performance metrics under the Palm distribution of the base station process and prove the consistency of their empirical estimators. We then prove that, for a typical base station, communication and sensing metrics are associated, so that operating regimes favourable to one functionality tend to benefit the other. The proof combines the monotonicity of the queue and filter recursions in the configuration of interferers with the Harris--FKG inequality for Poisson functionals. We also show that the excess association ratio of the two interference levels decays as the inverse of the network density. Simulations based on a ray-tracing channel model corroborate the analysis.
\end{abstract}


\section{Introduction}
\label{sec:intro}
Integrated sensing and communication (ISAC) is recognised as a core enabler of sixth-generation (6G) networks \cite{de2021convergent}. By integrating data transmission and environmental sensing within a shared infrastructure, ISAC promises significant improvements in both spectral and hardware efficiency \cite{liu2022survey}. However, realising this dual functionality in dense, large-scale deployments introduces substantial challenges, particularly in accurately modelling and evaluating network-wide performance. The principal difficulty lies in capturing the coupled spatio-temporal interactions among base stations (BSs), user equipments (UEs) and sensing objects (SOs).

In this work, we address this challenge by developing a dynamic stochastic geometry (SG) framework. BSs are modelled as static points drawn from a homogeneous Poisson point process (HPPP) on the plane, while UEs and SOs exhibit localised mobility around their serving BS, reflecting intra-cell motion. A common waveform is assumed for communication and sensing, with full spectrum reuse across BSs. Consequently, each BS generates interference that simultaneously affects communication links and sensing measurements, inducing a fundamental statistical coupling between the two functionalities. Multipath propagation is explicitly captured through a spatially distributed fading field. 

Communication performance is characterised by a downlink queue per UE, whose service rate is determined by the instantaneous signal-to-interference-plus-noise ratio (SINR) via the Shannon capacity \cite{blaszczyszyn2019performance}. Sensing performance is modelled using a Kalman filter (KF) whose measurement noise covariance depends on the instantaneous sensing SINR, which extends the classical KF, in which the noise covariance is exogenous \cite{Simon-book}. System-level performance metrics are defined under the Palm distribution of the BS point process, both for the time-averaged interference levels, SINRs and for the queue length and the KF error covariance.

Within this framework, we prove that communication and sensing metrics are associated at the typical BS. The proof exploits the fact that adding an interfering BS can only increase the interference perceived by both functionalities, which in turn degrades both the queue and the filter. Combined with the Harris--FKG inequality for Poisson functionals \cite{fortuin1971correlation,last2018lectures}, this monotonicity yields association without any assumption on the distribution of the fading. We further show that the excess association ratio of the two interference levels decays as the inverse of the BS density because the aggregate interference concentrates around its mean. Simulations based on a channel model derived from ray-tracing (RT) data corroborate the analysis. These findings indicate that communication and sensing performance can be jointly inferred and exploited, which provides a theoretical foundation for cross-functional control strategies in ISAC-enabled cellular networks.

\subsection*{Related work}
Prior work on ISAC~\cite{de2021convergent} has addressed waveform design~\cite{koivunen2024multicarrier,stratidakis2025near}, resource allocation~\cite{pulkkinen2024partially,vindas2025dynamic} and spectrum reuse~\cite{luong2021radio,liu2020joint}, as summarised in several surveys \cite{liu2022survey,luong2021radio}. A recurring theme is cross-operation-aware design, in which sensing outcomes inform transmission policies and vice versa~\cite{liu2022survey,olson2023coverage}. Yet a fundamental question remains open: how are communication and sensing performance coupled at the system level and how can this coupling be formally characterised? To address this question, we adopt SG tools~\cite{ baccelli:hal-02460214,andrews2011tractable}, which provide the analytical tractability needed to model large-scale spatial interactions rigorously.

\subsection*{Contributions}
\begin{itemize}
    \item We propose an SG framework for large-scale ISAC networks, define network-level performance metrics and prove the consistency of their empirical estimators.
    \item We prove that communication and sensing metrics are associated at the typical BS. We also characterise how the association of the interference levels depends on the network density.
    \item We validate the results through simulations based on a channel model derived from RT data.
\end{itemize}

\section{System Model}\label{sec:model}
All random elements below are defined on a single probability space $(\Omega,\mathcal F,\PP)$. Their joint law is fixed by the independence relations of Assumption~\ref{ass:indep}. Time is slotted, $t\in\NN =\{1,2,\dots\}$.

\subsection{Geometry and mobility}
\begin{figure}[t]
\centering
\includegraphics[width=0.8\columnwidth,trim=4 4 4 19,clip]{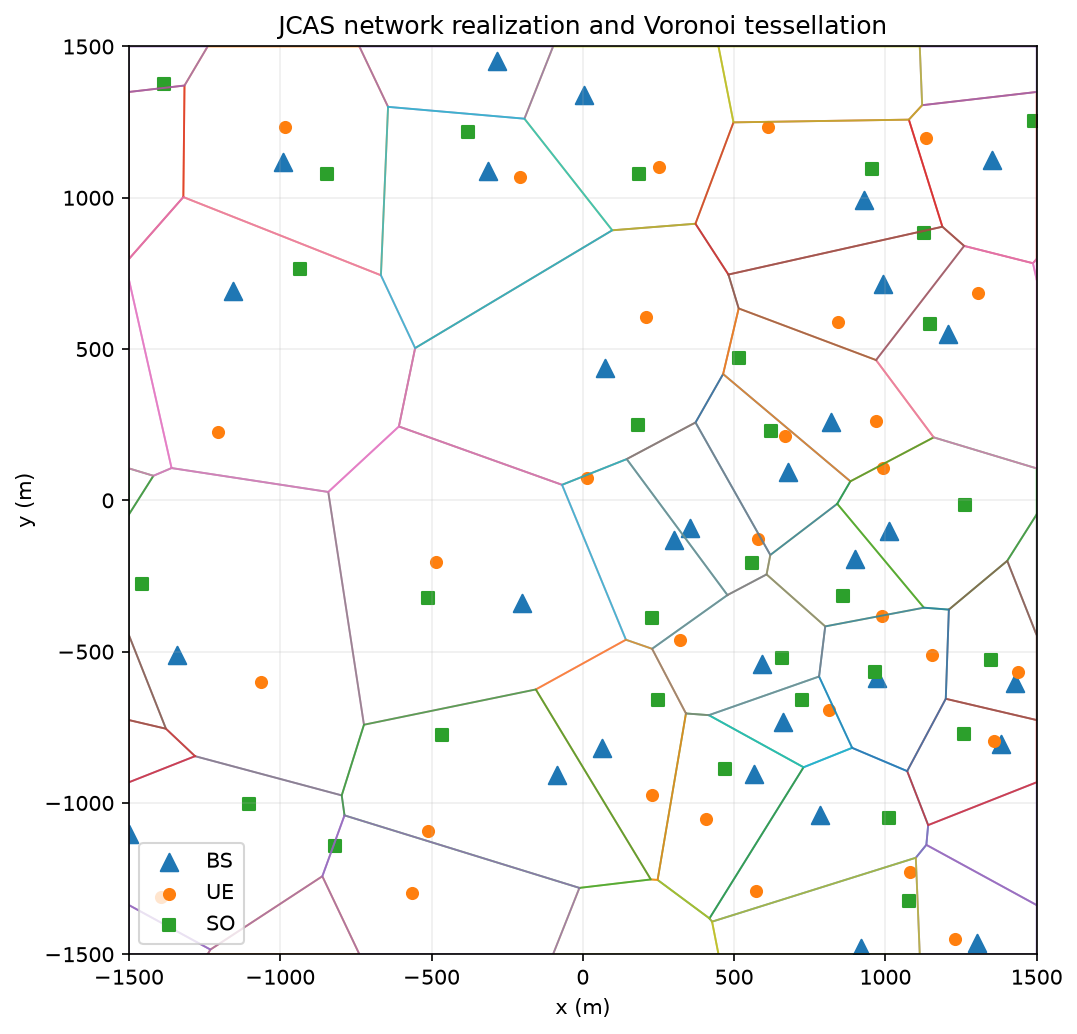}
\caption{One realisation of the network on a $9\,\mathrm{km}^2$ area: BSs (triangles) and their Voronoi tessellation, UEs (circles) and SOs (squares).}
\label{fig:network}
\end{figure}

The BS locations form an HPPP $X\subseteq\RR^2$ of intensity $\lambda>0$, enumerated $X=\{x_1,x_2,\ldots\}$ by increasing Euclidean norm. Each BS serves one UE and tracks one SO, both in captive motion: each particle remains associated with the same BS over the entire time horizon, so handovers are neglected. Figure \ref{fig:network} shows one realisation of the network. Generalising to finitely many UEs and SOs per BS is straightforward at the cost of heavier indexing. The UE and SO positions are driving processes translated by the BS location: $U_i(t)=x_i+\tilde U_i(t)$ and $S_i(t)=x_i+\tilde S_i(t)$. The UE driving processes $\tilde U_i$ are i.i.d.\ copies of a stationary ergodic discrete-time Markov process on $\RR^2$. The SO driving processes follow the Gauss--Markov recursion
\begin{equation}\label{eq:gm}
\tilde S_i(t)=G\tilde S_i(t-1)+w_i(t),
\end{equation}
where $G$ has spectral radius $\rho(G)<1$, the motion noises $w_i(t)\sim\mathcal N(0,Q)$ are i.i.d.\ with $Q$ positive definite and $\tilde S_i(0)$ is drawn from the stationary law of~\eqref{eq:gm}.

\subsection{Channel model}\label{sec:channel}
The power gain from a BS at $x\in\RR^2$ to a receiver at $y\in\RR^2$ in slot $t$ is
\begin{equation}\label{eq:power}
p(x,y;t)=\ell(|x-y|)\,F(x,y;t),
\end{equation}
where $\ell:[0,\infty)\to(0,1]$ is a non-increasing path loss, e.g.\ $\ell(r)=(r\vee1)^{-\alpha}$ with $\alpha>2$. Small-scale fading is modelled as a random field $\{F(x,y;t)\}$ of non-negative and mutually independent random variables, where $F(x,y;t)$ is the fading coefficient of the link from $x$ to $y$ in slot $t$. The distribution of $F(x,y;t)$ may depend on the link only through its length $|x-y|$, i.e., $F(x,y;t)=\phi(|x-y|,\varepsilon_{x,y}(t))$, where $\phi(r,\cdot)$ is the generalised inverse of the distribution function of the fading at distance $r$ with i.i.d.\ auxiliary random variables $\varepsilon_{x,y}(t) \sim\mathcal U(0,1)$.
\begin{remark}
Inter-node distances in cellular networks far exceed the channel
coherence distance~\cite{clarke1968statistical},
so fading coefficients across links are treated as statistically
independent~\cite{lu2021stochastic}. This assumption can be relaxed
under MIMO, where antenna spacing may fall below the coherence
distance~\cite{sanguinetti2019toward}. The shared
waveform and the assumption of a stationary target during the pilot interval
imply the same fading distribution for both functionalities.
\end{remark}

We write $\bar m(r)=\EE[p(x,y;t)]$, $r=|x-y|$, for the mean power profile.

\subsection{Interference and SINR}
Although orthogonal time-based operation suppresses self-interference, inter-cell interference persists due to full spectrum reuse. For a finite or countable set $Z\subseteq\RR^2$ and $y\in\RR^2$, the aggregate interference at $y$ is $\xi(y,Z;t)=\sum_{z\in Z}p(z,y;t)$. For BS $i$, the communication interference is measured at the UE and the sensing interference at the BS, which acts as a monostatic radar receiver:
\begin{align}
    \xi^\cm_i(t)&=\xi(U_i(t),X\setminus\{x_i\};t),\\
    \xi^\sn_i(t)&=\xi(x_i,X\setminus\{x_i\};t).
\end{align}
The sensing interference is dominated by the BS-to-BS illumination paths, whose power typically exceeds that of the echo returns from external scatterers. With transmit powers normalised to one, the SINRs are
\begin{align}
\SINR^\cm_i(t)&=\frac{p(x_i,U_i(t);t)}{N^\cm+\xi^\cm_i(t)},\label{eq:sinrc}\\
\SINR^\sn_i(t)&=\frac{\beta\,p(x_i,S_i(t);t)^2}{N^\sn+\xi^\sn_i(t)},\label{eq:sinrs}
\end{align}
with noise powers $N^\cm,N^\sn\ge0$ and a radar correction factor $\beta>0$ (e.g.\ $\beta=4\pi\sigma_{\mathrm{RCS}}/\lambda_c^2$ in free-space propagation, for radar cross-section $\sigma_{\mathrm{RCS}}$ and wavelength $\lambda_c$). The square reflects the two-way propagation of the sensing path, whose power decays as $r^{-2\alpha}$, consistently with the doubling of the power-law decay discussed in~\cite[Section~2.1.5]{tse2005fundamentals}.

\subsection{Communication: downlink queues}
The communication process is modelled as a downlink queue attached to each UE \cite{blaszczyszyn2019performance}, whose service rate is given by the Shannon law $\mu_i(t)=\log_2(1+\SINR^\cm_i(t))$. The queue length evolves according to Lindley's recursion
\begin{equation}\label{eq:lindley}
W_i(t)=[W_i(t-1)+A_i(t)-D_i(t)]^+,\qquad W_i(0)=0,
\end{equation}
where $(A_i(t))_{t\in\NN}$ is i.i.d.\ with mean $\nu>0$ and $\EE[D_i(t)\,\vert\,\mu_i(t)]=\mu_i(t)$. The departures are constructed as $D_i(t)=f_D(\mu_i(t),\varepsilon^{\mathrm D}_i(t))$, where $\varepsilon^{\mathrm D}_i(t)\sim\mathcal U(0,1)$ is an auxiliary random variable and $f_D(\mu,\cdot)$ is the generalised inverse of the conditional distribution function of $D_i(t)$ given $\mu_i(t)=\mu$. We assume that this conditional distribution is stochastically non-decreasing in the service rate, as is the case for deterministic or Poisson departures, hence, $f_D(\cdot,v)$ is non-decreasing for every $v$. Owing to the spatial variability of the SINR, the stability condition is not uniform throughout the network, so stable and unstable queues may coexist.

\subsection{Sensing: Kalman filter}
For every $i$, a KF estimates the position of SO $i$ through a state estimate $\hat S_i(t)$, an error covariance $\Sigma_i(t)$, measurements $Y_i(t)$ and a Kalman gain $K_i(t)$. Since the measurement noise covariance depends on the SINR, $\Sigma_i(t)$ does not follow the classical Riccati recursion and is only an estimate of the true error covariance. The filter is initialised at $\hat S_i(0)=x_i$ and $\Sigma_i(0)=0$. The a priori predictions follow the Gauss--Markov dynamics
\begin{align}
    \hat S^-_i(t)&=G(\hat S_i(t-1)-x_i)+x_i,\\
    \Sigma^-_i(t)&=G\Sigma_i(t-1)G^\top+Q.
\end{align}
The measurements are $Y_i(t)=H(S_i(t)-x_i)+(2\SINR^\sn_i(t))^{-1/2}v_i(t)$, where $H$ is the observation matrix and the $v_i(t)\sim\mathcal N(0,I_2)$ are i.i.d., so that the measurement noise covariance is $R_i(t)=\frac12(\SINR^\sn_i(t))^{-1}I_2$. This noise covariance follows from the low-SINR regime of~\cite{olson2023coverage} and is consistent with~\cite{liu2022survey,nussbaum2023based}. Since $\SINR^\sn_i(t)$ is observable at the serving BS, the KF is directly implementable. With $\hat Y^-_i(t)=H(\hat S^-_i(t)-x_i)$, the update equations are
\begin{align}
    K_i(t)&=\Sigma^-_i(t)H^\top\bigl(H\Sigma^-_i(t)H^\top+R_i(t)\bigr)^{-1},\\[2pt]
    \Sigma_i(t)&=\bigl(I_2-K_i(t)H\bigr)\Sigma^-_i(t),\\[2pt]
    \hat S_i(t)&=\hat S^-_i(t)+K_i(t)\bigl(Y_i(t)-\hat Y^-_i(t)\bigr).
\end{align}
Since $\Sigma^-_i(t)\succeq Q$ is positive definite, the covariance update admits the equivalent information form~\cite{woodbury1950inverting}
\begin{equation}\label{eq:info}
    \Sigma_i(t)^{-1}=\Sigma^-_i(t)^{-1}+2\SINR^\sn_i(t)H^\top H.
\end{equation}

\begin{assumption}[Independence]\label{ass:indep}
The following random elements are mutually independent: the BS process $X$, the UE driving processes $\tilde U_i$, the initial SO positions $\tilde S_i(0)$, the motion noises $w_i(t)$, the measurement noises $v_i(t)$, the arrivals $A_i(t)$ and the departure auxiliary variables $\varepsilon^{\mathrm D}_i(t)$, for $i,t\in\NN$ and the fading auxiliary variables $\varepsilon_{x,y}(t)$, for $x,y\in\RR^2$ and $t\in\NN$.
\end{assumption}

\section{ISAC Performance Metrics}\label{sec:metrics}
Let $\PP^0$ and $\EE^0$ denote the Palm probability and expectation of $X$. Under $\PP^0$, the \emph{typical BS} is placed at the origin and indexed by $0$. By Slivnyak's theorem~\cite[Thm.~3.2.4]{baccelli:hal-02460214}, the remaining BSs $X^!_0=X\setminus\{0\}$ form an HPPP of intensity $\lambda$ and all other random elements keep the law of Assumption~\ref{ass:indep}.

For a process $z$ and $T\in\NN$, let $\ol z_T=\frac1T\sum_{t=1}^Tz(t)$ and $\ol z_\infty=\limsup_{T\to\infty}\ol z_T\in[0,\infty]$. The communication and sensing metrics of BS $i$ at horizon $T\in\NN\cup\{\infty\}$ are any of
\begin{align}
M^\cm_{i,T}&\in\bigl\{\ol{\xi}{}^\cm_{i,T},\ \ol{\SINR}{}^\cm_{i,T},\ \ol W_{i,T}\bigr\},\\
M^\sn_{i,T}&\in\bigl\{\ol{\xi}{}^\sn_{i,T},\ \ol{\SINR}{}^\sn_{i,T},\ \ol{\tr\Sigma}_{i,T}\bigr\}.
\end{align}
Conditionally on $X$, the fading is independent in time and the driving processes are stationary and mixing. Hence, the infinite-horizon metrics are limits of the finite-horizon ones, which exist a.s.\ in $[0,\infty]$ and depend on $X$ only: by the Birkhoff ergodic theorem~\cite{birkhoff1931proof} for the interference and the SINR; by Loynes' stationary solution~\cite{loynes1962stability} for the queue length, with $\ol W_{i,\infty}<\infty$ when $\nu$ is below the long-run service rate of BS $i$ and $\ol W_{i,\infty}=\infty$ when $\nu$ exceeds it; under the contraction conditions of~\cite{bougerol1993kalman} for $\tr\Sigma$.

For measurable $f,g:[0,\infty]\to[0,\infty)$, the \emph{ISAC metric} and the \emph{ISAC association metric} of the typical BS are
\begin{align}
J_{f,g}(T)&=\EE^0\bigl[f(M^\cm_{0,T})\,g(M^\sn_{0,T})\bigr],\label{eq:J}\\
\mathcal A_{f,g}(T)&=\frac{J_{f,g}(T)}{\EE^0[f(M^\cm_{0,T})]\,\EE^0[g(M^\sn_{0,T})]}.\label{eq:A}
\end{align}
With $f=\mathbf 1_{(\gamma_\cm,\infty]}$ and $g=\mathbf 1_{(\gamma_\sn,\infty]}$ applied to the SINRs, $J_{f,g}$ is the probability that the typical BS meets both SINR targets and $\mathcal A_{f,g}=\PP^0(\ol{\SINR}{}^\cm_{0} > \gamma_\cm \mid \ol{\SINR}{}^\sn_{0} > \gamma_\sn)/\PP^0(\ol{\SINR}{}^\cm_{0} > \gamma_\cm)$ measures how much meeting the sensing target raises the probability of meeting the communication target. With $f(x)=g(x)=x$, $\mathcal A_{f,g}=1+\Cov^0(M^\cm,M^\sn)/(\EE^0[M^\cm]\,\EE^0[M^\sn])$ is called the \emph{association ratio} and $\mathcal A_{f,g}-1$ the \emph{excess association ratio}.

The metrics are estimated from a finite network. Let $X_L=X\cap[-L/2,L/2]^2$ and
\begin{equation}\label{eq:estimator}
\langle fg\rangle_{L,T}=\frac{1}{|X_L|}\sum_{x_i\in X_L}f(M^\cm_{i,T})\,g(M^\sn_{i,T}),
\end{equation}
with $\langle f\rangle_{L,T}$ and $\langle g\rangle_{L,T}$ defined analogously. The estimator of~\eqref{eq:A} is $\hat{\mathcal A}_{L,T}=\langle fg\rangle_{L,T}/(\langle f\rangle_{L,T}\langle g\rangle_{L,T})$.

\begin{theorem}[Consistency]\label{th:consistency}
Let $f,g$ be bounded and measurable.
(a) For every $T\in\NN$, a.s., $\langle fg\rangle_{L,T}\to J_{f,g}(T)$ as $L\to\infty$ and $\hat{\mathcal A}_{L,T}\to\mathcal A_{f,g}(T)$ whenever $\EE^0[f(M^\cm_{0,T})]\,\EE^0[g(M^\sn_{0,T})]>0$.
(b) If, $\PP^0$-a.s., $M^\bullet_{0,T}\to M^\bullet_{0,\infty}$ as $T\to\infty$ for $\bullet\in\{\cm,\sn\}$ and the limits are continuity points of $f$ and $g$, respectively, then, a.s., $\lim_{L}\lim_{T}\langle fg\rangle_{L,T}=J_{f,g}(\infty)=\lim_TJ_{f,g}(T)$ and likewise for $\mathcal A_{f,g}$.
\end{theorem}

\begin{IEEEproof}
\emph{Marked representation.} Only the joint law of the network matters, so we may realise it as follows. Mark each BS $x_i$ with $\mathsf M_i=(\tilde U_i,\tilde S_i,v_i,A_i,\varepsilon^{\mathrm D}_i,\varepsilon^\cm_i,\varepsilon^\sn_i,\varepsilon^{\mathrm U}_i,\varepsilon^{\mathrm B}_i)$, where each entry is indexed by time: $\varepsilon^\cm_i(t)$ and $\varepsilon^\sn_i(t)$ are the auxiliary variables of the serving link $x_i\to U_i(t)$ and of the echo link $x_i\to S_i(t)$ and $\varepsilon^{\mathrm U}_i(t)=(\varepsilon^{\mathrm U}_{i,k}(t))_{k\in\NN}$ and $\varepsilon^{\mathrm B}_i(t)=(\varepsilon^{\mathrm B}_{i,k}(t))_{k\in\NN}$ collect those of the links from the $k$-th closest other BS to UE $i$ and to BS $i$, respectively. Each directed link thus enters exactly one mark, that of its receiver, so the fading is independent across links, as Assumption~\ref{ass:indep} requires. The marked process $\mathcal X=\{(x_i,\mathsf M_i)\}$ is a stationary Poisson process with i.i.d.\ marks, hence mixing and ergodic under the spatial shifts $\theta_x$~\cite[Ch.~8]{baccelli:hal-02460214}. Every quantity of cell $i$ is a fixed measurable function of the configuration seen from $x_i$, i.e., $M^\bullet_{i,T}=h^\bullet_T(\theta_{x_i}\mathcal X)$ for $\bullet\in\{\cm,\sn\}$, with $h^\bullet_\infty=\limsup_Th^\bullet_T$.
\emph{(a)} The ergodic theorem for Palm expectations~\cite[Ch.~8]{baccelli:hal-02460214}, applied to the bounded functional $F=(f\circ h^\cm_T)(g\circ h^\sn_T)$, gives $(\lambda L^2)^{-1}\sum_{x_i\in X_L}F(\theta_{x_i}\mathcal X)\to\EE^0[F]=J_{f,g}(T)$ a.s. With $F\equiv1$, it yields $|X_L|/(\lambda L^2)\to1$. Dividing numerator and denominator of \eqref{eq:estimator} by $(\lambda L^2)$ leads to $\langle fg\rangle_{L,T}\to J_{f,g}(T)$. The same argument applies to $\langle f\rangle_{L,T}$ and $\langle g\rangle_{L,T}$ and the ratio converges by continuity.
\emph{(b)} Let $B$ be the set of configurations on which $h^\bullet_T\to h^\bullet_\infty$ fails or the limit is a discontinuity point of $f$ or $g$. By assumption, $\PP^0(B)=0$. By the Campbell--Mecke formula~\cite[Thm.~6.1.29]{baccelli:hal-02460214}, $\EE[\sum_{x_i\in X}\mathbf 1_B(\theta_{x_i}\mathcal X)]=\lambda\int_{\RR^2}\PP^0(B)\,\dd x=0$, so a.s.\ no BS lies in $B$. Hence, for every $L$, each of the finitely many terms of~\eqref{eq:estimator} converges as $T\to\infty$, so that $\lim_T\langle fg\rangle_{L,T}=\langle fg\rangle_{L,\infty}$. Part~(a), applied to $h^\bullet_\infty$, gives $\langle fg\rangle_{L,\infty}\to J_{f,g}(\infty)$ and $J_{f,g}(T)\to J_{f,g}(\infty)$ by dominated convergence. The ratio follows as in~(a).
\end{IEEEproof}

\section{Association}\label{sec:assoc}

\begin{definition}
    Two random variables $M_1$ and $M_2$ with values in $[0,\infty]$ are \emph{associated}~\cite{fortuin1971correlation} if, for every pair of non-decreasing functions $f,g:[0,\infty]\to[0,\infty)$,
    \begin{equation}\label{eq:exp_association}
        \EE[f(M_1)g(M_2)] \ge \EE[f(M_1)]\,\EE[g(M_2)].
    \end{equation}
\end{definition}

\emph{Representation under $\PP^0$.}
Under $\PP^0$, $U_0(t)=\tilde U_0(t)$ and $S_0(t)=\tilde S_0(t)$ and the randomness of the typical cell splits into three groups: the \emph{communication data} $Z^\cm=(\tilde U_0,A_0,\varepsilon^{\mathrm D}_0,\varepsilon^\cm_0)$, where $\varepsilon^\cm_0(t)$ is the auxiliary variable of the serving link $0\to U_0(t)$; the \emph{sensing data} $Z^\sn=(\tilde S_0,v_0,\varepsilon^\sn_0)$, where $\varepsilon^\sn_0(t)$ is that of the echo link $0\to S_0(t)$; the \emph{interferers} $\Xt=\{(x,\mathsf m_x):x\in X^!_0\}$, where the mark $\mathsf m_x=(\varepsilon^\cm_x(t),\varepsilon^\sn_x(t))_{t\in\NN}$ collects the auxiliary variables of the links from $x$ to the typical UE and to the typical BS. Here, $A_0=(A_0(t))_{t\in\NN}$ and similarly for the other sequences. Attaching these auxiliary variables to the interferers does not alter the joint law, since they are i.i.d.\ and independent of all atoms of $X^!_0$. By Slivnyak's theorem, the marking theorem~\cite[Thm.~5.6]{last2018lectures} and Assumption~\ref{ass:indep}, $\Xt$ is a Poisson process on $\RR^2\times\mathbb M$, with $\mathbb M$ the mark space. $\Xt$, $Z^\cm$ and $Z^\sn$ are independent. Moreover,
\begin{align}
\xi^\cm_0(t)&=\sum_{(x,\mathsf m_x)\in\Xt}\ell(|x-U_0(t)|)\,\phi\bigl(|x-U_0(t)|,\varepsilon^\cm_x(t)\bigr),\nonumber\\
\xi^\sn_0(t)&=\sum_{(x,\mathsf m_x)\in\Xt}\ell(|x|)\,\phi\bigl(|x|,\varepsilon^\sn_x(t)\bigr),\label{eq:shotnoise}
\end{align}
so $\xi^\cm_0$ is a function of $(\Xt,Z^\cm)$ and $\xi^\sn_0$ of $\Xt$ alone. 

\begin{lemma}[Pathwise monotonicity]\label{lem:mono}
Fix $Z^\cm$, $Z^\sn$ and marked configurations $\varphi\subseteq\varphi'$. Denote by a prime the quantities obtained with $\Xt=\varphi'$ in place of $\Xt=\varphi$. For every $t\in\NN$ and $\bullet\in\{\cm,\sn\}$:
(i) $\xi^\bullet_0(t)\le\xi'^\bullet_0(t)$ and $\SINR^\bullet_0(t)\ge\SINR'^\bullet_0(t)$;
(ii) $W_0(t)\le W'_0(t)$;
(iii) $\Sigma_0(t)\preceq\Sigma'_0(t)$ in the Loewner order, hence $\tr\Sigma_0(t)\le\tr\Sigma'_0(t)$.
Consequently, for every $T\in\NN\cup\{\infty\}$, each of $\ol\xi{}^\cm_{0,T}$, $-\ol{\SINR}{}^\cm_{0,T}$ and $\ol W_{0,T}$ is a non-decreasing function of $\Xt$ for fixed $Z^\cm$ and each of $\ol\xi{}^\sn_{0,T}$, $-\ol{\SINR}{}^\sn_{0,T}$ and $\ol{\tr\Sigma}_{0,T}$ is a non-decreasing function of $\Xt$ for fixed $Z^\sn$.
\end{lemma}

\begin{IEEEproof}
(i) By~\eqref{eq:shotnoise}, $\xi'^\bullet_0(t)-\xi^\bullet_0(t)$ is the sum of the non-negative contributions of the interferers in $\varphi'\setminus\varphi$ and the numerators of the SINRs involve only the typical BS, which is not a point of $\Xt$.
(ii) By (i), $\mu'_0(t)\le\mu_0(t)$, hence $D'_0(t)\le D_0(t)$, since $f_D(\cdot,v)$ is non-decreasing. As $(w,d)\mapsto[w+A_0(t)-d]^+$ is non-decreasing in $w$ and non-increasing in $d$, so $W_0(t-1)\le W'_0(t-1)$ implies $W_0(t)\le W'_0(t)$. Induction on $t$ from $W_0(0)=W'_0(0)=0$ concludes. 
(iii) Assume $\Sigma_0(t-1)\preceq\Sigma'_0(t-1)$. Then $\Sigma^-_0(t)\preceq\Sigma'^-_0(t)$, both matrices being positive definite since $Q\succ0$. Since matrix inversion reverses the Loewner order on positive definite matrices and $\SINR'^\sn_0(t)\le\SINR^\sn_0(t)$ by (i), the information form~\eqref{eq:info} gives
\begin{align*}
\Sigma_0(t)^{-1}&=\Sigma^-_0(t)^{-1}+2\SINR^\sn_0(t)H^\top H\\
&\succeq\Sigma'^-_0(t)^{-1}+2\SINR'^\sn_0(t)H^\top H=\Sigma'_0(t)^{-1}.
\end{align*}
Inverting once more yields $\Sigma_0(t)\preceq\Sigma'_0(t)$ and induction on $t$ from $\Sigma_0(0)=\Sigma'_0(0)=0$ concludes. The trace is Loewner-monotone. Finally, time averages and $\limsup$ preserve these inequalities.
\end{IEEEproof}

\begin{theorem}[Association]\label{th:association_pairs}
    Under $\PP^0$, for every $T\in\NN\cup\{\infty\}$, each of the pairs $(\ol\xi{}^\cm_{0,T},\ol\xi{}^\sn_{0,T})$, $(\ol{\SINR}{}^\cm_{0,T},\ol{\SINR}{}^\sn_{0,T})$ and $(\ol W_{0,T},\ol{\tr\Sigma}_{0,T})$ is associated. In particular, $\mathcal A_{f,g}(T)\ge1$ for these pairs and every non-decreasing $f,g$ with finite positive marginal expectations.
\end{theorem}

\begin{IEEEproof}
We detail the pair $(\ol W,\ol{\tr\Sigma})$, dropping the subscripts $0,T$. It suffices to consider bounded $f$ and $g$, since the general case follows by monotone convergence applied to $f\wedge n$ and $g\wedge n$. Let $Z=(Z^\cm,Z^\sn)$. By Lemma~\ref{lem:mono}, conditionally on $Z$, both $f(\ol W)$ and $g(\ol{\tr\Sigma})$ are non-decreasing functionals of the Poisson process $\Xt$, which is independent of $Z$. The Harris--FKG inequality for Poisson functionals~\cite[Thm.~20.4]{last2018lectures} therefore gives
\begin{align*}
\EE^0\bigl[f(\ol W)\,g(\ol{\tr\Sigma})\bigm|Z\bigr]
&\ge\EE^0\bigl[f(\ol W)\bigm|Z^\cm\bigr]\,\EE^0\bigl[g(\ol{\tr\Sigma})\bigm|Z^\sn\bigr],
\end{align*}
where the conditional expectations on the right depend only on $Z^\cm$ and on $Z^\sn$, respectively, because $\ol W$ is a function of $(\Xt,Z^\cm)$ and $\ol{\tr\Sigma}$ of $(\Xt,Z^\sn)$. Since $Z^\cm$ and $Z^\sn$ are independent, taking expectations yields~\eqref{eq:exp_association}. The interference pair is treated in the same way. For the SINR pair, $f(\ol{\SINR}{}^\cm)$ and $g(\ol{\SINR}{}^\sn)$ are both non-increasing functionals of $\Xt$, to which the Harris--FKG inequality applies equally.
\end{IEEEproof}

Association yields lower bounds on joint events in terms of marginal probabilities. For thresholds $\gamma_W,\gamma_\Sigma\ge0$, the choice $f=\mathbf 1_{(\gamma_W,\infty]}$ and $g=\mathbf 1_{(\gamma_\Sigma,\infty]}$ gives, dropping the subscripts $0,T$,
\begin{align}
\PP^0\bigl(\ol W>\gamma_W,\,\ol{\tr\Sigma}>\gamma_\Sigma\bigr)
\ge\PP^0\bigl(\ol W>\gamma_W\bigr)\,\PP^0\bigl(\ol{\tr\Sigma}>\gamma_\Sigma\bigr),
\end{align}
which lower-bounds the probability of joint queue congestion and sensing degradation. Similarly, the probability that the typical BS meets both SINR targets is at least the product of the probabilities of meeting each of them. Hence, marginal coverage probabilities, available in closed form for classical SG models~\cite{andrews2011tractable} or measurable separately for each functionality, give certified lower bounds on the joint ISAC coverage.

\begin{proposition}[Dilution law]\label{prop:dilution}
Let $\mathcal I_k=\int_{\RR^2}\bar m(|x|)^k\,\dd x$ and assume that $\bar m>0$, $\mathcal I_1,\mathcal I_2<\infty$. For every $T\in\NN$, the association ratio of the interference pair, i.e.~\eqref{eq:A} with $M^\bullet=\ol\xi{}^\bullet_{0,T}$ and $f(x)=g(x)=x$, is
\begin{equation}\label{eq:dilution}
\mathcal A=1+\frac{\kappa}{\lambda},\qquad
\kappa=\frac{\EE^0\bigl[(\bar m\star\bar m)(\tilde U_0(1))\bigr]}{\mathcal I_1^2},
\end{equation}
where $(\bar m\star\bar m)(u)=\int_{\RR^2}\bar m(|x-u|)\,\bar m(|x|)\,\dd x$. Moreover, $0<\kappa\le\mathcal I_2/\mathcal I_1^2$, with equality iff $\tilde U_0(1)\equiv0$.
\end{proposition}

\begin{IEEEproof}
We drop the subscripts $0,T$. By Tonelli, $\ol\xi{}^\cm=\sum_{(x,\mathsf m_x)\in\Xt}a_x$ and $\ol\xi{}^\sn=\sum_{(x,\mathsf m_x)\in\Xt}b_x$, where $a_x$ and $b_x$ are the time averages over $t\le T$ of the summands in~\eqref{eq:shotnoise}. Averaging over the marks,
\begin{align*}
\bar a(x)&:=\EE^0[a_x\mid Z^\cm]=\tfrac1T\textstyle\sum_{t=1}^T\bar m(|x-U_0(t)|),\\
\bar b(x)&:=\EE^0[b_x]=\bar m(|x|),
\end{align*}
and $\EE^0[a_xb_x\mid Z^\cm]=\bar a(x)\,\bar b(x)$, since $\varepsilon^\cm_x$ and $\varepsilon^\sn_x$ are independent. Conditionally on $Z^\cm$, the trajectory $U_0(1),\dots,U_0(T)$ is fixed. Hence Campbell's formula, the bivariate Mecke equation~\cite[Thm.~4.4]{last2018lectures}, the translation invariance of the Lebesgue measure and Tonelli's theorem give
\begin{align*}
\EE^0[\ol\xi{}^\cm\mid Z^\cm]
&=\lambda\textstyle\int\bar a\,\dd x
=\tfrac{\lambda}{T}\sum_{t=1}^T\int\bar m(|x-U_0(t)|)\,\dd x\\
&=\lambda\mathcal I_1=\EE^0[\ol\xi{}^\sn],\\
\EE^0[\ol\xi{}^\cm\ol\xi{}^\sn\mid Z^\cm]
&=\lambda\textstyle\int\bar a\,\bar b\,\dd x+\lambda\int\bar a\,\dd x\cdot\lambda\int\bar b\,\dd x\\
&=\lambda\textstyle\int\bar a(x)\,\bar m(|x|)\,\dd x+\lambda^2\mathcal I_1^2.
\end{align*}
Taking expectations and using the stationarity of $\tilde U_0$,
\begin{align*}
\Cov^0(\ol\xi{}^\cm,\ol\xi{}^\sn)&=\tfrac{\lambda}{T}\textstyle\sum_{t=1}^T\EE^0\bigl[(\bar m\star\bar m)(\tilde U_0(t))\bigr]\\
&=\lambda\,\EE^0\bigl[(\bar m\star\bar m)(\tilde U_0(1))\bigr],
\end{align*}
and dividing by $\EE^0[\ol\xi{}^\cm]\,\EE^0[\ol\xi{}^\sn]=(\lambda\mathcal I_1)^2$ yields~\eqref{eq:dilution}. By Cauchy--Schwarz and translation invariance, $(\bar m\star\bar m)(u)\le\mathcal I_2$, with equality iff $u=0$, which gives the bounds on $\kappa$. Note that $\kappa>0$ since $\bar m>0$.
\end{IEEEproof}

The constant $\kappa$ is the autocorrelation of the mean power profile $\bar m$ at the UE displacement, normalised by $\mathcal I_1^2$. It quantifies the contribution of interferers that are strong at both the UE and its BS and is maximised when the UE coincides with the BS. Since the covariance of the two interferences grows linearly with $\lambda$, whereas the product of their means grows quadratically, the excess association ratio $\kappa/\lambda$ vanishes as the network densifies. We refer to this effect as \emph{dilution of influence}.

\section{Numerical Results}\label{sec:sim}
To assess the framework under realistic propagation, the simulations use a stochastic multipath channel model fitted to RT data of a suburban area of Oulu, Finland, at carrier frequency $f_c=1$\,GHz with isotropic antennas~\cite{Pyhtila2023}. For a link of length $d$, the model draws a distance-dependent number of paths with random excess delays, gains and phases\footnote{The RT channel model has been designed to address ISAC scenarios within the SNS JU INSTINCT 6G project. Details will be available in~\cite{instinctD43}.}. BSs are drawn as a HPPP on a torus to exclude boundary effects, each with one captive UE and one captive SO. The parameters are listed in Table~\ref{tab:params}. Metrics are time-averaged over $T$ slots and averaged over all BSs as in~\eqref{eq:estimator} and $\hat{\mathcal A}_{L,T}$ is computed with $f(x)=g(x)=x$. All results can be reproduced with the public simulator~\cite{jcasSimulator}. 
\begin{table}[t]
\caption{Simulation parameters}
\label{tab:params}
\centering\footnotesize
\begin{tabular}{@{}ll@{}}
\hline
Carrier, bandwidth & $f_c=1$\,GHz, $B=20$\,MHz \\
Transmit power, noise PSD & $46$\,dBm, $-174$\,dBm/Hz \\
Radar cross-section & $\sigma_{\mathrm{RCS}}=1$\,m$^2$ \\
BS density, torus area & $\lambda=3.33$, $9\,\mathrm{km}^2$ ($\approx30$ BSs) \\
Horizon & $T=500$ \\
SO model~\eqref{eq:gm}, KF & $\rho(G)=0.99$, $\rho(Q)=1.5$, $H=I_2$ \\
UE displacement & same as SO model \\
Arrivals & $\nu=1$ bit/s/Hz per slot \\
\hline
\end{tabular}
\end{table}

\begin{figure*}[t]
\centering
\includegraphics[width=0.3\textwidth,trim=5 8 5 20,clip]{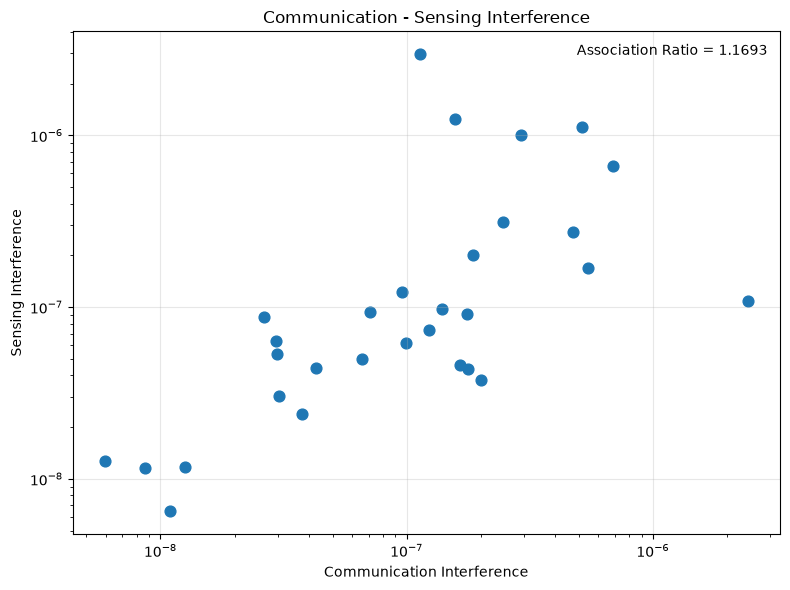}\hspace{0.03\textwidth}
\includegraphics[width=0.3\textwidth,trim=5 7 5 20,clip]{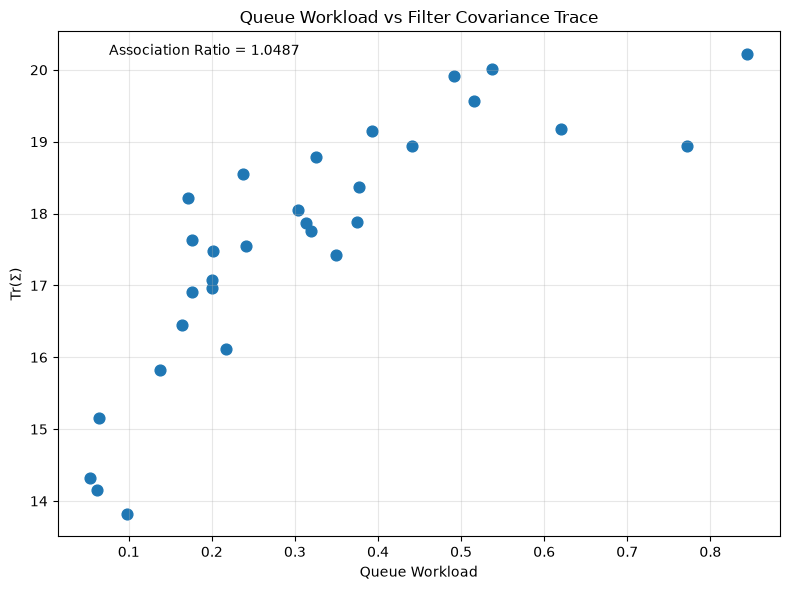}\hspace{0.03\textwidth}
\includegraphics[width=0.3\textwidth,trim=9 11 9 25,clip]{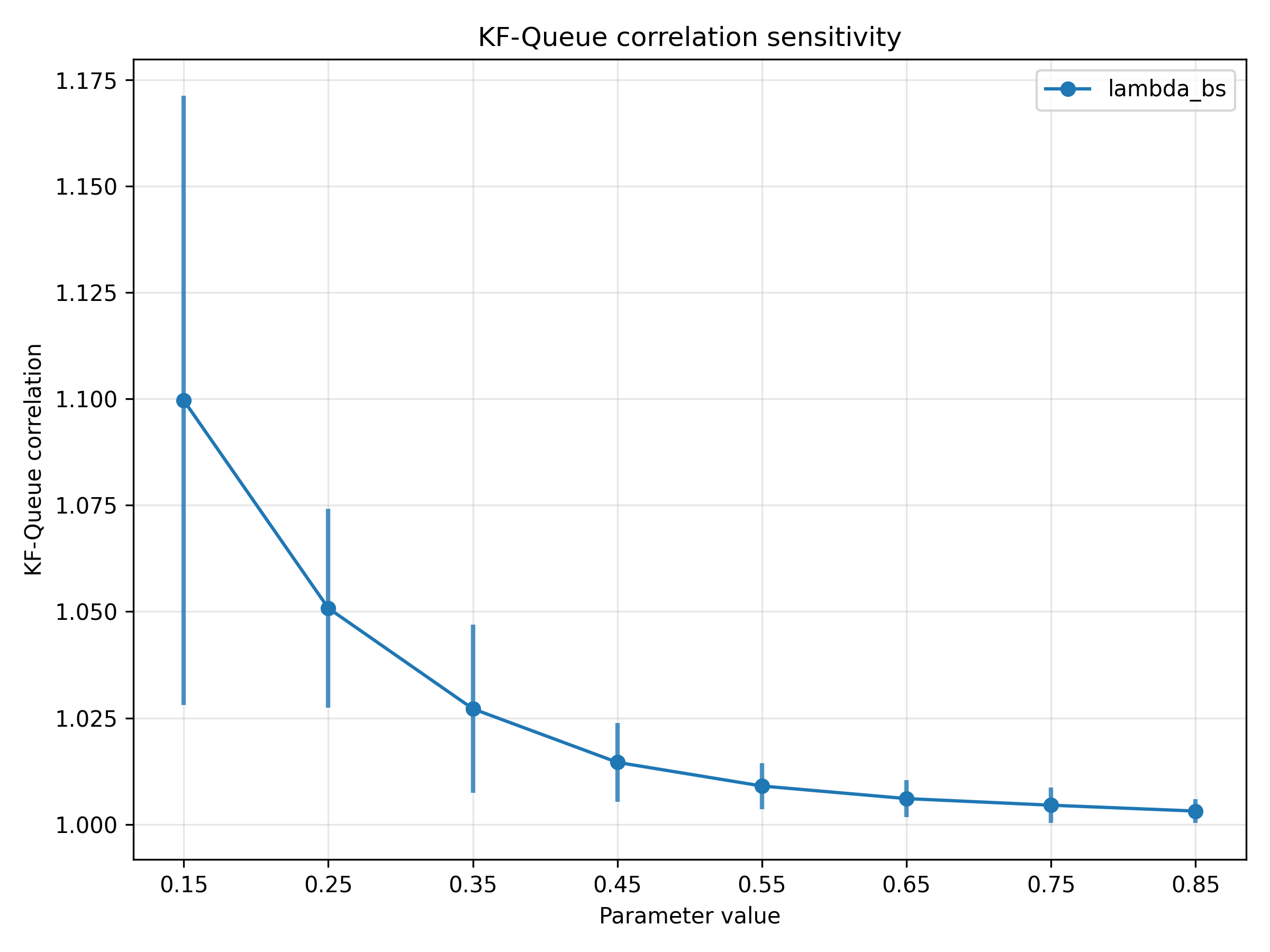}
\caption{From left to right: (a) $(\ol\xi{}^\cm_{i,T},\ol\xi{}^\sn_{i,T})$ and (b) $(\ol W_{i,T},\ol{\tr\Sigma}_{i,T})$ for every BS of one realisation; (c) $\hat{\mathcal A}_{L,T}$ of $(\ol W,\ol{\tr\Sigma})$ versus $\lambda$ ($100\,\mathrm{km}^2$, sweep settings $0.15$--$0.85$), mean and standard deviation over 100 realisations.}
\label{fig:results}
\end{figure*}

\subsection{Results}
Figures~\ref{fig:results}(a) and~(b) display the time-averaged metrics of every BS in one realisation of the network. The association ratio equals $1.17$ for the interference pair and $1.05$ for the pair formed by the queue length and the trace of the KF error covariance. Both values exceed one, in agreement with Theorem~\ref{th:association_pairs}: strong communication interference comes with strong sensing interference and congested queues with poorly tracked SOs. 

Figure~\ref{fig:results}(c) shows the association ratio of $(\ol W,\ol{\tr\Sigma})$ as a function of $\lambda$. It remains above one and decreases towards one as the network densifies, consistently with the dilution law of Proposition~\ref{prop:dilution}, suggesting an inverse-density scaling law similar to that of the interference pair.

\section{Conclusion}
We developed a stochastic-geometry framework to characterise the dependence structure of ISAC networks under full spectrum reuse, together with empirical performance metrics whose consistency we established. Our main result shows that, at the typical BS, the interference levels, SINRs, queue lengths and KF error covariances are associated across the two functionalities. The proof relies on the monotonicity of the model with respect to the interferer configuration and the Harris--FKG inequality for Poisson functionals, without imposing any assumption on the fading distribution. We further show that the excess association ratio of the interference pair decays inversely with the network density. Simulations based on a channel model derived from RT data corroborate these findings, suggesting that information from one functionality can be leveraged to improve the performance of the other.

\section*{Acknowledgment}
The authors are grateful to Fran\c{c}ois Baccelli for his continuous guidance and thank Joonas Kokkoniemi, Pekka Sangi, Juha Pyhtil\"a and Visa Tapio for providing the RT channel model and related parameters. This work was partly supported by the SNS JU Horizon Europe Project under Grant Agreement No. 101139161
(INSTINCT).

\bibliographystyle{IEEEtran}
\bibliography{biblio}

\begin{thebibliography}{10}
\providecommand{\url}[1]{#1}
\csname url@samestyle\endcsname
\providecommand{\newblock}{\relax}
\providecommand{\bibinfo}[2]{#2}
\providecommand{\BIBentrySTDinterwordspacing}{\spaceskip=0pt\relax}
\providecommand{\BIBentryALTinterwordstretchfactor}{4}
\providecommand{\BIBentryALTinterwordspacing}{\spaceskip=\fontdimen2\font plus
\BIBentryALTinterwordstretchfactor\fontdimen3\font minus \fontdimen4\font\relax}
\providecommand{\BIBforeignlanguage}[2]{{%
\expandafter\ifx\csname l@#1\endcsname\relax
\typeout{** WARNING: IEEEtran.bst: No hyphenation pattern has been}%
\typeout{** loaded for the language `#1'. Using the pattern for}%
\typeout{** the default language instead.}%
\else
\language=\csname l@#1\endcsname
\fi
#2}}
\providecommand{\BIBdecl}{\relax}
\BIBdecl

\bibitem{de2021convergent}
C.~De~Lima, D.~Belot, R.~Berkvens, A.~Bourdoux, D.~Dardari, M.~Guillaud, M.~Isomursu, E.-S. Lohan, Y.~Miao, A.~N. Barreto \emph{et~al.}, ``Convergent communication, sensing and localization in 6g systems: An overview of technologies, opportunities and challenges,'' \emph{IEEE Access}, vol.~9, pp. 26\,902--26\,925, 2021.

\bibitem{liu2022survey}
A.~Liu, Z.~Huang, M.~Li, Y.~Wan, W.~Li, T.~X. Han, C.~Liu, R.~Du, D.~K.~P. Tan, J.~Lu \emph{et~al.}, ``A survey on fundamental limits of integrated sensing and communication,'' \emph{IEEE Communications Surveys \& Tutorials}, vol.~24, no.~2, pp. 994--1034, 2022.

\bibitem{blaszczyszyn2019performance}
B.~B{\l}aszczyszyn and M.~K. Karray, ``Performance analysis of cellular networks with opportunistic scheduling using queueing theory and stochastic geometry,'' \emph{IEEE Transactions on Wireless Communications}, vol.~18, no.~12, pp. 5952--5966, 2019.

\bibitem{Simon-book}
D.~Simon, \emph{Optimal state estimation: Kalman, H infinity, and nonlinear approaches}.\hskip 1em plus 0.5em minus 0.4em\relax John Wiley \& Sons, 2006.

\bibitem{fortuin1971correlation}
C.~M. Fortuin, P.~W. Kasteleyn, and J.~Ginibre, ``Correlation inequalities on some partially ordered sets,'' \emph{Communications in Mathematical Physics}, vol.~22, no.~2, pp. 89--103, 1971.

\bibitem{last2018lectures}
G.~Last and M.~Penrose, \emph{Lectures on the Poisson process}.\hskip 1em plus 0.5em minus 0.4em\relax Cambridge University Press, 2018, vol.~7.

\bibitem{koivunen2024multicarrier}
V.~Koivunen, M.~F. Keskin, H.~Wymeersch, M.~Valkama, and N.~Gonzalez-Prelcic, ``Multicarrier isac: Advances in waveform design, signal processing, and learning under nonidealities,'' \emph{IEEE Signal Processing Magazine}, vol.~41, no.~5, pp. 17--30, 2024.

\bibitem{stratidakis2025near}
G.~Stratidakis, S.~Droulias, and A.~Alexiou, ``Near-field hierarchical localization for integrated sensing and communication,'' \emph{IEEE Access}, vol.~13, pp. 17\,766--17\,776, 2025.

\bibitem{pulkkinen2024partially}
P.~Pulkkinen and V.~Koivunen, ``Partially observable model-based learning for isac resource allocation,'' in \emph{ICASSP 2024-2024 IEEE International Conference on Acoustics, Speech and Signal Processing (ICASSP)}.\hskip 1em plus 0.5em minus 0.4em\relax IEEE, 2024, pp. 12\,996--13\,000.

\bibitem{vindas2025dynamic}
Y.~Vindas and M.~Guillaud, ``Dynamic channel charting: Integrating online sample selection with continual learning for streaming csi data,'' in \emph{ICC 2025-IEEE International Conference on Communications}.\hskip 1em plus 0.5em minus 0.4em\relax IEEE, 2025, pp. 3791--3796.

\bibitem{luong2021radio}
N.~C. Luong, X.~Lu, D.~T. Hoang, D.~Niyato, and D.~I. Kim, ``Radio resource management in joint radar and communication: A comprehensive survey,'' \emph{IEEE Communications Surveys \& Tutorials}, vol.~23, no.~2, pp. 780--814, 2021.

\bibitem{liu2020joint}
F.~Liu, C.~Masouros, A.~P. Petropulu, H.~Griffiths, and L.~Hanzo, ``Joint radar and communication design: Applications, state-of-the-art, and the road ahead,'' \emph{IEEE Transactions on Communications}, vol.~68, no.~6, pp. 3834--3862, 2020.

\bibitem{olson2023coverage}
N.~R. Olson, J.~G. Andrews, and R.~W. Heath, ``Coverage and rate of joint communication and parameter estimation in wireless networks,'' \emph{IEEE Transactions on Information Theory}, vol.~70, no.~1, pp. 206--243, 2023.

\bibitem{baccelli:hal-02460214}
\BIBentryALTinterwordspacing
F.~Baccelli, B.~Blaszczyszyn, and M.~K. Karray, \emph{{Random Measures, Point Processes, and Stochastic Geometry}}.\hskip 1em plus 0.5em minus 0.4em\relax {Inria}, Jul. 2024. [Online]. Available: \url{https://inria.hal.science/hal-02460214}
\BIBentrySTDinterwordspacing

\bibitem{andrews2011tractable}
J.~G. Andrews, F.~Baccelli, and R.~K. Ganti, ``A tractable approach to coverage and rate in cellular networks,'' \emph{IEEE Transactions on communications}, vol.~59, no.~11, pp. 3122--3134, 2011.

\bibitem{clarke1968statistical}
R.~H. Clarke, ``A statistical theory of mobile-radio reception,'' \emph{Bell system technical journal}, vol.~47, no.~6, pp. 957--1000, 1968.

\bibitem{lu2021stochastic}
X.~Lu, M.~Salehi, M.~Haenggi, E.~Hossain, and H.~Jiang, ``Stochastic geometry analysis of spatial-temporal performance in wireless networks: A tutorial,'' \emph{IEEE Communications Surveys \& Tutorials}, vol.~23, no.~4, pp. 2753--2801, 2021.

\bibitem{sanguinetti2019toward}
L.~Sanguinetti, E.~Bj{\"o}rnson, and J.~Hoydis, ``Toward massive mimo 2.0: Understanding spatial correlation, interference suppression, and pilot contamination,'' \emph{IEEE Transactions on Communications}, vol.~68, no.~1, pp. 232--257, 2019.

\bibitem{tse2005fundamentals}
D.~Tse and P.~Viswanath, \emph{Fundamentals of wireless communication}.\hskip 1em plus 0.5em minus 0.4em\relax Cambridge New York, NY, 2005, vol.~1.

\bibitem{nussbaum2023based}
A.~Nussbaum, W.~D. Blair, B.~Keel, and U.~Ramachandran, ``Em-based radar signal processing and tracking of maneuvering targets,'' in \emph{2023 IEEE International Radar Conference (RADAR)}.\hskip 1em plus 0.5em minus 0.4em\relax IEEE, 2023, pp. 1--6.

\bibitem{woodbury1950inverting}
M.~A. Woodbury, \emph{Inverting modified matrices}.\hskip 1em plus 0.5em minus 0.4em\relax Department of Statistics, Princeton University, 1950.

\bibitem{birkhoff1931proof}
G.~D. Birkhoff, ``Proof of the ergodic theorem,'' \emph{Proceedings of the National Academy of Sciences}, vol.~17, no.~12, pp. 656--660, 1931.

\bibitem{loynes1962stability}
R.~M. Loynes, ``The stability of a queue with non-independent inter-arrival and service times,'' in \emph{Mathematical proceedings of the cambridge philosophical society}, vol.~58, no.~3.\hskip 1em plus 0.5em minus 0.4em\relax Cambridge University Press, 1962, pp. 497--520.

\bibitem{bougerol1993kalman}
P.~Bougerol, ``Kalman filtering with random coefficients and contractions,'' \emph{SIAM Journal on Control and Optimization}, vol.~31, no.~4, pp. 942--959, 1993.

\bibitem{Pyhtila2023}
J.~Pyhtilae, J.~Kokkoniemi, P.~Sangi, N.~Vaara, and M.~Juntti, ``Ray tracing based radio channel modelling applied to ris,'' in \emph{WSA \& SCC 2023; 26th International ITG Workshop on Smart Antennas and 13th Conference on Systems, Communications, and Coding}.\hskip 1em plus 0.5em minus 0.4em\relax VDE, 2023, pp. 1--6.

\bibitem{instinctD43}
\BIBentryALTinterwordspacing
{INSTINCT Consortium}, ``{INSTINCT} deliverable {4.3},'' to appear. [Online]. Available: \url{https://cordis.europa.eu/project/id/101139161/results}
\BIBentrySTDinterwordspacing

\bibitem{jcasSimulator}
\BIBentryALTinterwordspacing
E.~Mengoli, ``{INSTINCT} {JCAS} simulator,'' GitHub repository. [Online]. Available: \url{https://github.com/emanuelemengoli/instinct-jcas-simulator}
\BIBentrySTDinterwordspacing

\end{thebibliography}

\end{document}